\documentclass[12pt,american]{article}


\usepackage{graphicx}
\usepackage{amsmath, amssymb, amsthm}
\usepackage{natbib}
\usepackage{filecontents}
\usepackage{ascmac}
\usepackage{subfigure}

\usepackage[mathscr]{euscript} 
\usepackage{bbm}

\usepackage{enumitem}

\usepackage{changes}
\newcommand{\stkout}[1]{\ifmmode\text{\sout{\ensuremath{#1}}}\else\sout{#1}\fi}

\setdeletedmarkup{\stkout{#1}}
\setlist[enumerate]{itemsep=1.5pt,topsep=1.5pt}
\setlist[itemize]{itemsep=1.5pt,topsep=1.5pt}

\renewcommand{\leq}{\leqslant}
\renewcommand{\geq}{\geqslant}

\renewcommand{\phi}{\varphi}
\renewcommand{\epsilon}{\varepsilon}

\usepackage{mdwlist}

\usepackage[citecolor=blue, colorlinks=true, linkcolor=blue]{hyperref}

\usepackage{caption}
\usepackage{floatrow}

\usepackage{algorithmic}
\usepackage{algorithm}

\usepackage[left=1.in, right=1.in, top=.5in, bottom=.5in, includehead, includefoot]{geometry}

\newcommand{\RR}{\mathbbm R}

\theoremstyle{plain}
\newtheorem{lemma}{Lemma}
\newtheorem{proposition}{Proposition}

\theoremstyle{definition}
\newtheorem{definition}{Definition}
\newtheorem{remark}{Remark}
\newtheorem{assumption}{Assumption}

\usepackage[multiple]{footmisc}

\usepackage{enumitem}

\usepackage{mathtools}
\usepackage{authblk}

\title{Pecuniary Externality, Ideology and Sphere of Influence\thanks{We would like to thank Han Huynh who worked on this project for many years, and our colleagues who gave us useful comments, in particular, Pierpaolo Battigalli, Kohei Kawamura,   So Kubota, Takashi Kunimoto, Jong-Wha Lee, Massimo Morelli, Marco Ottaviani, Daisuke Oyama, Danny Quah, Satoru Takahashi, Serene Tan, and Romain Wacziarg. This work was supported by JSPS KAKENHI Grant Numbers JP22H00849, JP21K13263 and Waseda University's Special Research Project 2021C010 and 2023C247. Corresponding author: Tomoo Kikuchi. Email: \texttt{tomookikuchi@waseda.jp}}
	
}

\author[a]{Tomoo Kikuchi}
\author[b]{Shuige Liu}
\author[c]{Lien Pham}

\affil[a,c]{\small Graduate School of Asia-Pacific Studies, Waseda University}
\affil[b]{\small Department of Decision Sciences, Bocconi University}

\begin{document}

	\maketitle

\thispagestyle{empty}

	\vspace{-.5cm}	

	\begin{abstract} 
	
\noindent We build a game-theoretic model to formalize \citeauthor{kindleberger1996world}'s \citeyearpar{kindleberger1996world} idea of the public good of leadership. Two superpowers use ideology to compete for influence by forming clubs whose members benefit more when their orientations are closer to the club’s. Pecuniary externalities create complementarity among non-superpowers. Their collective agency forces superpower to compromise by choosing an orientation not aligned with its own. We find that superpowers compromise most when expanding their clubs but less as members become more dependent. Moreover, increased member endowments do not always enlarge the club; disproportionate growth by a few can instead contract it.

	\vspace{.5ex}
		
	\noindent\textbf{Keywords:} superpowers; club goods; deterrence; coalition formation;  subgame perfect  equilibrium
		
	\noindent\textbf{JEL\ Classification:} C7; D6;  D7; F5 
	
	\end{abstract}
	
\clearpage

\pagenumbering{arabic}
	
\begin{quote}\small
``Economic primacy, at its best, involves less dominance or hegemony than the public good of leadership of the world economy, not ordering others to behave as the leader directs, but pointing the way and convincing others of the desirability of following.'' \citep[][p.13]{kindleberger1996world}
\end{quote}

\section{Introduction}


China has significantly increased its global influence over the past several decades across economic, political, technological, and military spheres.
For example, China's global outreach can be seen in the increasing number of countries that have signed a Memorandum of Understanding (MOU) with the Chinese government to participate in the Belt and Road Initiative (BRI). Since its launch in 2013, the BRI membership has grown rapidly reaching 146 countries as of May 2025. In particular, China has expanded the membership by investing in authoritarian regimes that many Western investors avoided.\footnote{Source: The Democracy Index, The Economist Intelligence Unit.}

This paper aims to analyze the underlying forces when superpowers compete for their sphere of influence focusing on the role of ideology. To this end, we develop a game-theoretic model to analyze how two superpowers compete for a sphere of influence by sequentially choosing  an ideological orientation of their clubs that can, to a varying degree, accommodate countries with different ideological orientations, who decide whether to join a club or not.

Club goods are exclusive to members but not rivalrous as increasing members does not diminish the utility from the club goods. However, we assume that members benefit more from a club good when their own orientations are closer to that of the club. Therefore, superpowers may strategically compromise in choosing a club orientation that does not align with its own to attract more members.

Pecuniary externalities of joining a club naturally lead to complementary in decision making of non-superpowers. We then have an interaction between non-cooperative players, who compete to form clubs, and cooperative players, who coordinate to join the clubs. This interaction provides insights into a mechanism through which non-superpowers have collective agency to influence the sphere of influence of superpowers.

By modeling superpowers who make their club desirable to non-superpowers who have collective agency, we formalize 
\citeauthor{kindleberger1996world}'s \citeyearpar{kindleberger1996world} idea of leadership as  ``pointing the way and convincing others of the desirability of following.'' This is in contrast to recent works addressing  spheres of influence competition from the viewpoint of coercion or bilateral alignment \citep{jackson2020understanding, camboni2024spheres, clayton2025putting}.
 
The pecuniary  externalities arise in our model as the cost for providing a club good is shared among members, i.e., the more countries join the club, the less each member's cost share becomes. For example, the countries, who signed a MOU for the BRI, receive investments from China
mainly in the energy, transportation and technology sectors that have pecuniary externalities enhancing trade and connectivity with China and other countries. Therefore, BRI projects are usually financed by China's government financial institutions and host country governments who share the costs.

\begin{figure}[ht!]
	\centering
	\includegraphics[width=0.5\columnwidth]{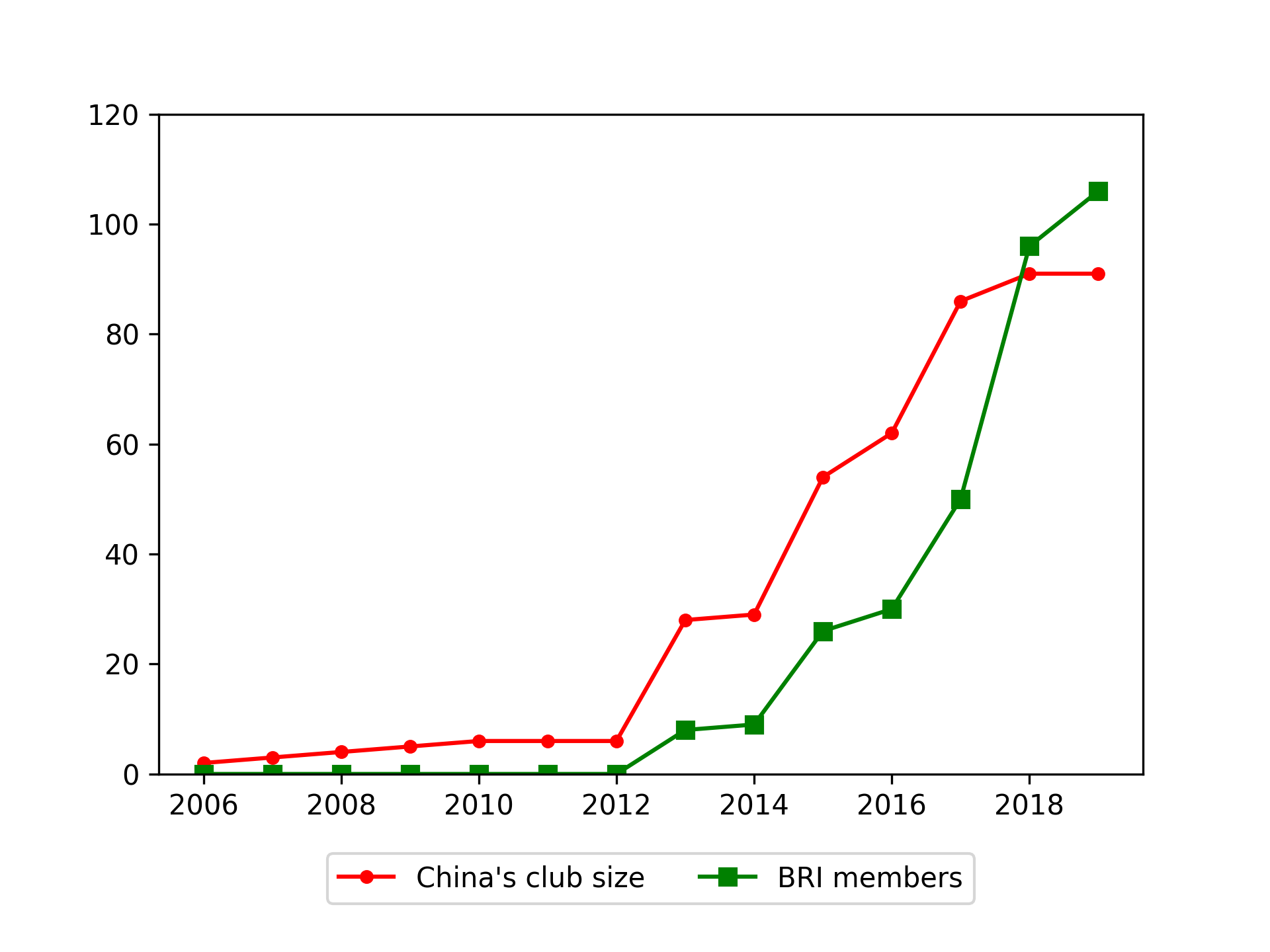}
	\caption{Simulation result and BRI membership}
	\label{fig:bri}
\end{figure}

Figure \ref{fig:bri} shows the simulation results of our games played every year where the US and China compete to attract members among 144 countries. China's club size depicts the number of countries that choose to join China's club instead of the US's club or staying out of any club. We can observe that our model captures the dynamics of expanding BRI membership fairly well.

How countries derive utility from a club in our model is inspired by how utility is modeled in relation to the location of a good in \cite{hotelling1929stability}.
Non-superpowers in our model resemble citizens who pick a community that best matches their preference. This principle of mobile citizens ``voting with their feet'' goes back to the idea by \cite{tiebout1956pure}. 
Our model builds on foundational concepts and frameworks on clubs and memberships \citep{tiebout1956pure, pauly1970cores, sorenson1978private, cornes1996theory}. However, our extension allows for inter-club competition by strategic positioning, where superpowers adjust club orientations to attract members under competitive pressure---a dimension unexplored in the traditional club good literature.

Our paper is inspired by the literature on great powers in international relations. For example, 
we share with \cite{hirschman1980national} the insight that trade dependencies are a source of geopolitical power. 
\cite{lake1984beneath} studies international economic structures where states play different structural roles depending on their relative size and productivity. \cite{barrett1994self} builds a game-theoretic model and examines self-enforcing international treaties, offering valuable insights into how clubs maintain stability through incentive compatibility.
\citeauthor{mearsheimer2003tragedy}'s \citeyearpar{mearsheimer2003tragedy} theory on great power competition assumes that great powers always act according to their own self-interest.
We contribute to the literature by studying the strategic agency and collective actions of non-superpowers that shape superpower competition and their spheres of influence. 

The rest of the paper is organized as follows. Section \ref{sec:two-stage} introduces our model and presents our solution concepts.  Section \ref{sec:special} analyzes comparative statics. Section \ref{sec:US-China} presents simulation results of the the US-China competition. Section \ref{sec:conclusion} concludes. 
The remainder of proofs can be found in the appendix

\section{The two-stage sequential game}\label{sec:two-stage}

This section presents the formal model, outlining the strategic interactions between superpowers and how they are influenced by collective choices of non-superpowers. Section \ref{sec:framework} introduces the general framework of the two-stage sequential game. Section \ref{sec:supermodular} introduces a coalition-formation process that describes the behavior of non-superpowers.
Given the behavior of non-superpowers, Section \ref{sec:sub-game} derives the choices of superpowers in  a subgame perfect equilibrium (SPE). 

\subsection{The general framework}\label{sec:framework}

The world consists of countries, among which there are two superpowers $H = \{a, b\}$ and a finite set of non-superpowers $i\in I$.  Each country $j \in H \cup I$ has an orientation in a finite metric space $L$ and an endowment $\omega_j\in\RR^+$. 
Let a club be a set of countries who choose to benefit from the club good provided by a superpower. 
The superpowers compete for a sphere of influence, measured by their club membership.
Given the choices of superpowers, non-superpowers collectively decide whether to join a club.  
The club good is non-rivalrous but excludable, benefiting only its members without being depleted. Each non-superpower can join at most one club. 
Club members (including the superpower) share the fixed cost of maintaining their club. We assume that a superpower cannot become a member of another superpower's club. 

The two superpowers play a two-stage sequential Stackelberg game.  
In stage one, the status-quo power $a$ decides whether to open a club or not and, if so, the orientation of the club.
In stage two, the challenger $b$ makes its decision. 
The choices of non-superpowers are represented by a vector  $c=(c_{i})_{i \in I} \in  \{a,b,0\}^{I}$ where $c_i=a$ (or $b$)  means that $i$ chooses to join $a$ (or $b$)'s club and $c_i=0$ means that $i$ chooses to stay out of any club.
The payoff of a non-superpower from joining a club depends on three factors:
the distance between its own orientation and  the club's orientation, its dependence on
the club's superpower, and the club members. Formally, given the two superpower's choices $\ell_a, \ell_b\in L$, the utility of non-superpower $i \in I$ is given by 
\begin{equation}\label{SEC}
	\tag{NU}
	\underbrace{u_{i}(\ell_{a},\ell_{b}, c)}_{\text{utility}} = 
	\begin{cases}
		\underbrace{g_{i,c_{i}}\times(1-d_{i,\ell_{c_{i}}})}_{\text{benefit}} - \underbrace{\rho_{i,c_{i}}(I_{c_{i}}(c))}_{\text{cost}} & \text{ if }  c_{i}\in H\\
		0 & \text{ if }c_{i} = 0.
	\end{cases}
\end{equation}
Here,  $g_{i,c_{i}}$ is a measure of $i$'s dependency on $c_{i}$.  Given $i$'s choice $c_i$, when $c_i=a$ or $b$, $d_{i, \ell_{c_{i}}}$ is the distance between $i$'s orientation and $\ell_{c_{i}}$. $I_{c_{i}}(c) := \{j \in I: c_{j} = c_{i}\}$ is the set of non-superpowers who choose to join $c_{i}$'s club. 
Finally, $\rho_{i,c_{i}}(I_{c_{i}})$ determines the cost for non-superpower $i$, which is strictly decreasing in $I_{c_{i}}$ and in $\omega_j$ for all $j\in I_{c_i}\cup \{c_i\}\setminus \{i\}$, but strictly increasing in $\omega_i$.  
The utility function \eqref{SEC} shows that a non-superpower derives greater utility from a club when its dependency on the superpower is higher or when its orientation is closer to the club's orientation. 

Given $\ell_{a},\ell_{b}\in L$, and the choice vector $c$ of non-superpowers, 
the utility of superpower $e \in H$  is given by 
\begin{equation}\label{UYT}
	\tag{SU}
\underbrace{u_{e}(\ell_{e}, c)}_{\text{utility}} = 
	\begin{cases}
			\underbrace{1-d_{e,\ell_{e}}}_{\text{benefit}}	 - \underbrace{\rho_{e}(I_{e}(c))}_{\text{cost}} & \text{ if }  \ell_{e} \in L\\
		0 & \text{ if no club}.
	\end{cases}
\end{equation}
When a superpower perfectly aligns the club's orientation with its own orientation, the benefit $1-d_{e,\ell_{e}}$ is at the maximum. However, to attract more members and lower its cost, it may choose an orientation away from its own. In doing so, the superpower makes a strategic {\em compromise}—adjusting the club's orientation to appeal to more non-superpowers. The utility function \eqref{UYT} captures this trade-off: a superpower chooses a club's orientation further from its own, only when the increasing membership offsets the cost through shared contributions.

\subsection{Coalition formation}\label{sec:supermodular}

This section introduces a coalition-formation process that describes the behavior of non-superpowers. 
The process allows sequential deviations until no subset of non-superpowers can deviate and be better off.  Given the choices of superpowers, the game among non-superpowers is essentially a voting game in the sense of \cite{peleg2002game}. 
Moreover, it leads, by definition, to a Nash, a strong Nash, and a coalition-proof Nash equilibrium. 

In stage one, given $a$'s choice $\ell_{a}$, there is a group $I_{1}$ of non-superpowers  who join $a$'s club,  as they obtain a positive payoff. This might attract another group $I_{2}$ of countries  to join the club, which might attract yet another group $I_{3}$ to join the club, and so on. This is because the club is non-rivalrous, and the payoff increases when new members join and share the cost. The process stops at step $m$, when no outside group can benefit from joining the club. 
Given $a$'s choice $\ell_{a}$ and a coalition-formation process $I_{1},...,I_{m}$, we define $I^{*}(\ell_{a}) = \cup_{k=1}^{m}I_{k}$ as $a$'s sphere of influence.
The formal algorithm is introduced in Appendix \ref{sec:algorithm}, where we show that  the process is order-independent,
implying that $I^{*}(\ell_{a})$ is well defined. 
The resulting  $a$'s club is given by $I^*(\ell_a) \cup \{a\}$. The process leads to a core outcome, such that no subgroup of non-superpowers can deviate in a way that makes every member strictly better off, guaranteeing stability in the coalition structure.

In stage two, given $\ell_{a}$ and the club $I^{*}(\ell_{a})$, if $b$ chooses $\ell_{b}$ for its club, non-superpowers need to adjust their choices. This can be described as a process of coordination, in which countries in and outside $I^{*}(\ell_{a})$ evaluate if they can achieve a strictly higher payoff in $b$'s club. The formal algorithm is given in Appendix \ref{sec:algorithm}, where we show that  the coalition-formation process is order-independent. Here, as in stage one, there is no other subgroup that can improve every member's payoff by  changing the membership. 
The club formed by $a$ in stage one serves as a benchmark, based on which non-superpowers behave in stage two. 
Everyone who switches from $a$ to $b$'s club should improve its payoff. We denote the final clubs of $a$ and $b$ as $I_a(\ell_a,\ell_b)$ and $I_b(\ell_a,\ell_b)$ respectively.

Mathematically, the final clubs can be defined in one shot, by treating the game as a supermodular game in the sense of \cite{milgrom1990rationalizability} and by modifying \cite{topkis1979equilibrium}'s algorithm. 
In our viewpoint, however, it is reasonable to think of the two stages as two independent coalition-formation processes. 
When a status-quo power establishes a club, it chooses its club's orientation, knowing that it might be challenged by an emerging power. On the other hand, non-superpowers do not consider how well off they would be in the club of an emerging power, when they consider whether to join the status-quo power's club  in stage one. 
Only after an emerging power establishes a new club in stage two, non-superpowers decide whether to switch to the new club. 
In other words, we assume that in stage one,  non-superpowers do not take into account the choice of a superpower that might form a club in stage two. 

The assumption that non-superpowers consider their payoffs independently in stage one and in stage two naturally captures a bechmarking effect in a sequential game, where the payoff of each non-superpower in stage one serves as a reference point for evaluating $b$'s offer in stage two. 
Knowing this benchmarking effect,  $a$ strategically places its club to attract enough members to make it difficult for $b$ to form a viable alternative, potentially deterring its club formation altogether.

\subsection{Sub-game perfect equilibrium}\label{sec:sub-game}

The sequential game gives a first-mover advantage for the status-quo power $a$ and is also essential for capturing how a real-world  power takes an initiative to defend the status-quo in geopolitical competition rather than merely reacting to an emerging power. 
Given  $\ell_{a}$ and $\ell_{b}$, the choice of non-superpowers is represented by a vector $c^{\langle I_{a}(\ell_{a},\ell_{b}), I_{b}(\ell_{a},\ell_{b})\rangle}$, i.e., countries in $I_{a}(\ell_{a},\ell_{b})$ choose $a$, those in $I_{b}(\ell_{a},\ell_{b})$ choose $b$, and the remaining countries choose 0. 
The behavior of non-superpowers uniquely determines the payoffs for the superpowers for any club orientation choices. We can now rewrite \eqref{UYT} to express the payoff for each superpower $e\in H$ as $U_{e}(\ell_{a},\ell_{b}) = u_{e}(\ell_{e}, c^{\langle I_{a}(\ell_{a},\ell_{b}), I_{b}(\ell_{a},\ell_{b}) \rangle} )$. Given the choice of $a$, $b$ chooses a best response, whereas $a$ takes $b$'s response into account when it makes its choice. This backward induction leads to a SPE. Then, Kuhn's Theorem \citep{kuhn1953} guarantees that the game has at least one SPE. The following assumptions on the behavior of superpowers ensure uniqueness.

\begin{assumption}\label{assumption1}
	When the payoffs are the same for multiple orientations of the club, superpowers choose the closest one to home.
\end{assumption}

\begin{assumption}	\label{assumption2}
	When the payoff of forming a club is 0, superpowers do not establish a club.
\end{assumption}

Both assumptions break ties when superpowers are indifferent between choices.  
We obtain the following result.

\begin{lemma}\label{lemma_spe}
	Suppose that Assumptions \ref{assumption1} and \ref{assumption2} are satisfied and $d_{e,\ell}\neq d_{e,\ell^\prime}$ for $\ell\neq\ell^\prime$ where $\ell, \ell^\prime \in L$ and  $e\in H$.  The game has a unique subgame perfect  equilibrium.
\end{lemma}

The condition $d_{e,\ell}\neq d_{e,\ell^\prime}$ for $\ell\neq\ell^\prime$ means that for each superpower, a distinct orientation has a distinct distance from its own orientation. 
This condition is sufficient yet not necessary. 

\begin{remark}
	An infinitesimal change in  $g_{ie}$, $\omega_i$, or $\omega_e$ can lead to a sudden change in club size.\footnote{That is, there may exist $\hat{g}_{ie}$ such that $\limsup \limits_{g_{ie} \to \hat{g}_{ie}} \delta_e (g_{ie}) \geq \delta_e (\hat{g}_{ie}) + 2$ or $\liminf \limits_{g_{ie} \to \hat{g}_{ie}} \delta_e (g_{ie}) \leq \delta_e (\hat{g}_{ie}) - 2$. Similar formal statements can be stated for $\omega_i$ and $\omega_e$.}
\end{remark}

This insight follows directly from the coalition-formation process. When dependency on a superpower increases or cost share decreases to a point that a non-superpower is willing to join the club (jointly with some others), this decision can attract others to follow.

\subsection{Example}\label{sec:gametree}

We show that coalition formation weakens the first-mover advantage and induces compromise by the status-quo power to deter the emerging power.   
Consider the case where there are three non-superpowers located at $1/4, 1/2$ and $3/4$, the status-quo power $a$ is located at $0$, and the emerging power $b$ is located at $1$. 
The set of orientations is $\{0, 1/4, 1/2, 3/4, 1\}$, with distance
between any two orientations defined as their absolute difference.
The dependencies of non-superpowers on $a$ and $b$ are set to be $1/2$ and the endowments of all countries are set to be $1$. 

\begin{figure}[ht!]
	\centering
	\includegraphics[width=.8\columnwidth]{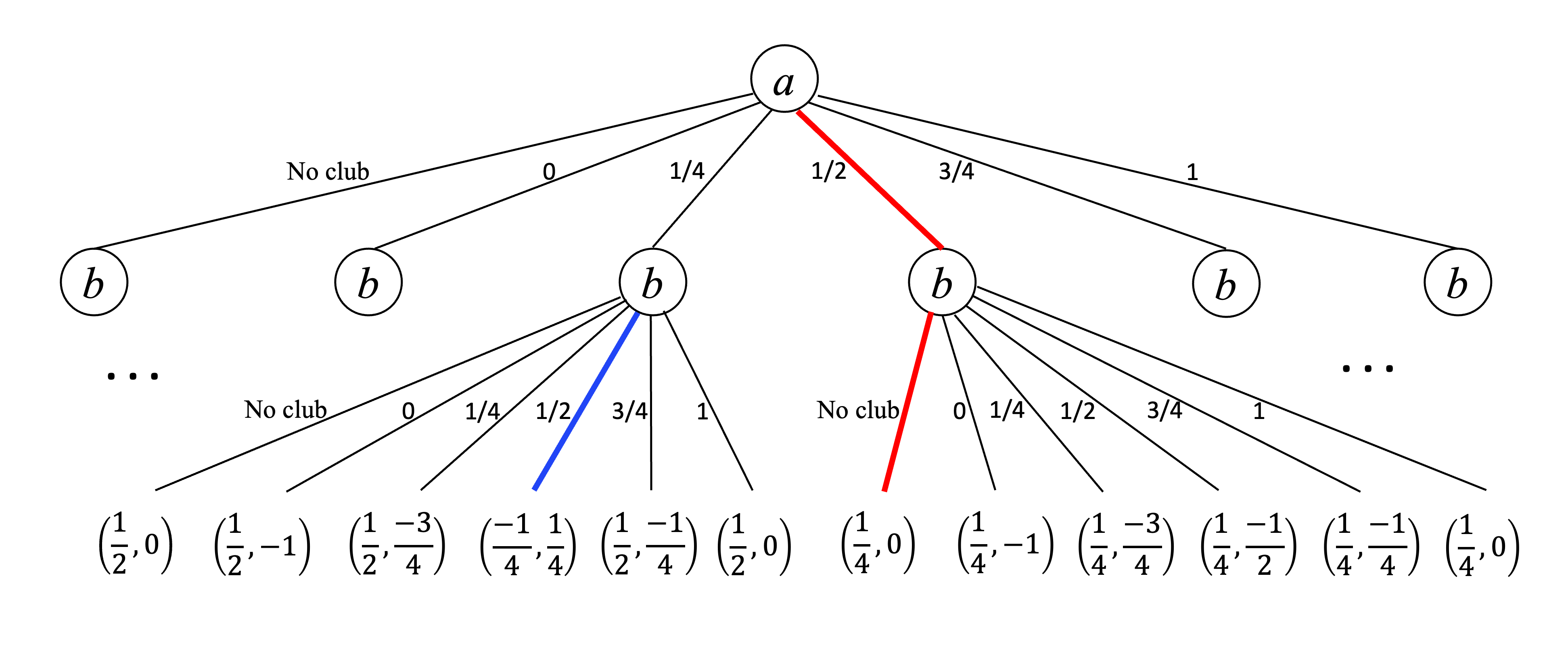}
	\caption{Coalition formation: Three non-superpowers}
	\label{fig:game_tree}
\end{figure}

Suppose $a$ chose $1/4$ so that all countries would join $a$'s club as they would get a positive payoff in stage one.
When non-superpowers did not coordinate their behavior, $b$ would decide not to open a club, since no one would move to $b$'s club \textit{alone}, no matter which orientation $b$ would choose for its club. 
On the other hand, when non-superpowers coordinate their behavior, if $b$ chooses $1/2$, the first round of the coalition-formation process assigns countries at $1/2$ and $3/4$ to $b$'s club, since by switching \textit{together} they gain strictly higher payoffs than staying in $a$'s club. 
Then, the country at $1/4$ gains a negative payoff if it remains in $a$'s club, 0 if it joins no club, and a positive payoff if it switches to $b$'s club. Hence, the second round of the coalition-formation process assigns this country to $b$'s club, and $b$ attracts everyone, as depicted in blue in Figure \ref{fig:game_tree}.
Hence, choosing $1/4$ is not optimal for $a$, as it gets a negative payoff. 
Therefore, when non-superpowers coordinate their behavior, $a$ chooses a more centrist orientation, $1/2$, in stage one to deter $b$, making everyone stay in $a$'s club, which is the SPE outcome depicted in red  in Figure \ref{fig:game_tree}. This shows that communication among non-superpowers weakens $a$'s first-mover advantage in a sense that it requires $a$ to compromise to deter $b$. 
Such collective action is a realistic feature of international relations, where
coordination among countries can significantly influence superpower competition. 

\section{Comparative statics}\label{sec:special}

We say that a superpower $e$ compromises more (less) when a superpower chooses on orientation so that $d_{e, \ell_e}$ takes a higher (lower) value.
Since $L$ is finite, for each superpower, there exists an orientation that is furthest away from its own orientation. Then, the distance from superpower $e$ to such an orientation $\bar{d_e} \coloneqq \max_{\ell \in L}d_{e, \ell}$ is the \textit{maximum compromise} the superpower may choose given $L$.

Next, let us define the \textit{robust club} of superpower $e$ given  $a$'s club orientation and $b$'s best response as the set of non-superpowers that would remain in $e$'s club if $e$ increases its compromise (up to the  maximum). 
\begin{definition}
	Let $\ell_b(\ell_a)$ be $b$'s best response to $\ell_a$.
	Given $\ell_a \in L$, the robust club of each superpower is:
	\[
	\underline{I}_a(\ell_a) \coloneqq 
	\bigcap_{\;\ell_a' : \, d_{a,\ell_a} \,\leq\, d_{a,\ell_a'} \,\leq\, \bar{d}_a}
	I_a(\ell_a', \ell_b(\ell_a'))
	\quad \text{ and } \quad
	\underline{I}_b(\ell_a) \coloneqq
	\bigcap_{\;\ell_b' : \, d_{b,\ell_b(\ell_a)} \,\leq\, d_{b,\ell_b'} \,\leq\, \bar{d}_b}
	I_b(\ell_a, \ell_b')
	.
	\]
\end{definition}
When a superpower compromises more to attract additional members, it reduces the utility of non-superpowers that are ideologically closer to itself. If such countries also have large endowments and therefore bear higher cost shares, they may leave the club when the superpower compromises further. Hence, the robust club primarily consists of countries that are highly dependent on the superpower in trade or that have relatively small endowments. In principle, the robust club of a superpower given $a$'s choice and $b$'s best response could be empty. However, most countries exhibit significant trade dependence on major powers, and the endowment gap between superpowers and other countries is substantial. If cost shares of members are proportional to their GDP, many non-superpowers would bear negligible costs when joining a club, implying that the robust club is likely to be non-empty.

\begin{proposition} \label{dependency}
	When robust club members become more dependent on the club's superpower, the superpower compromises weakly less.
\end{proposition}

Intuitively, when robust club members become more dependent on their club's superpower, their incentive to remain in the club strengthen even without additional compromise by the superpower. Countries outside the robust club, however, do not change their stance. As a result, for any given additional compromise, the resulting club size is no larger than before the increase in  dependency of members. Therefore, the superpower can maintain the same club size with less compromise.
Proposition \ref{dependency} thus implies that as  dependencies of countries on a superpower increase from low levels, the superpower initially compromises to attract members and form a club. Once its members become more dependent on it, however, the superpower compromises less.

In the following, we analyze how changing endowments of countries may affect the club size.
\begin{proposition} \label{endowment}
	Let $\omega, \omega' \in \mathbb{R}^n_{+}$ denote two endowment vectors.
	Fix any pair of orientations $(\ell_a, \ell_b) \in L^2$. For each $e \in \{a,b\}$, let $I_e$ and $I'_e$ denote its clubs under $\omega$ and $\omega'$, respectively. 
	
	\begin{enumerate}
		\item If $\omega'_i > \omega_i$ and 
		$\rho_{ie}(I_e;\omega')=\rho_{ie}(I_e;\omega)$ for all $i \in I_e \cup \{e\}$
		 while $\omega'_j = \omega_j$ for all $j \in I \cup \{-e\} \backslash I_e$, then $I'_e \supseteq I_e$. 
		That is, if the endowments of a superpower and its members increase but their cost shares remain unchanged, then the club weakly expands.
		\item If $\omega'_e > \omega_e$ and $\omega'_j = \omega_j$ for all $j \in I \cup \{-e\}$, then $I'_e \supseteq I_e$. That is, if the endowment of a superpower increases, then its club weakly expands.
	\end{enumerate}
\end{proposition}

Intuitively, the club becomes more attractive, allowing the superpower to retain all existing members and potentially gain new ones, when the endowments of the club's superpower and its members increase in a way such that their cost shares remain unchanged. When only the  endowment of the superpower increases, the cost of joining its club falls for all potential members, thus the club (weakly) expands. 

In the previous argument, we showed that when endowments of original club members increase in a ``balanced" way so that their cost shares do not change, the club can attract additional members given any pair of orientation choices by superpowers. However, this result may not hold when the endowment of only one (or a few) non-superpower grows disproportionately. While a higher endowment of one member reduces the cost for others, making the club more attractive, it  weakens that member's incentive to stay in the club as its cost share increases.
In this case, in the SPE outcome, the superpower may compromise to retain that member, potentially attracting others ideologically further away. Conversely, if the superpower cannot retain it, the member's departure may induce others to leave the club too. The result is summarized below. 

\begin{proposition} \label{endowment2}
	Consider a game where superpower $e$ forms a club $I_e$ in the SPE outcome. In general, for any $i \in I_e$, the club does not necessarily expand as $\omega_i$ increases.
\end{proposition}

\begin{proof}[Proof of Proposition \ref{endowment2}]
	We prove the proposition with a counterexample. Consider our example in Section \ref{sec:gametree}.  
	Suppose the endowment of the country at 1/2 becomes 4, while that of all others remains 1. Then, in the SPE outcome, $a$ chooses $\ell_a=1/4$ with the country at 1/4 as its member, and $b$ chooses $\ell_b = 3/4$ with the country at 3/4 as its member. The country at 1/2 chooses to stay out of any club given any choices of $a$ and $b$.
\end{proof}

\section{Simulation: The US-China competition}\label{sec:US-China}

To simulate the US-China competition, we use the trade ratio of countries with the US and China as a proxy for the dependency of non-superpowers on superpowers,\footnote{Source: UN Comtrade. Trade ratio with the US = [Export + Import with the US]/Total trade (in USD). The same for China.} 
the Democracy Index (DI) for the orientation of each country,
and GDP  for the endowment of countries.\footnote{Source: The World Bank and the OECD.} 
We assume that the cost share of country $j$ in club $e$ takes the form 
\begin{equation}
	\rho_{je} (I_{e}(c))= \frac{\omega_{j}}{\omega_{e} + \sum_{i \in I_{e}(c)} \omega_{i}}
\end{equation}
where $\omega_{j} >0$ is the GDP for each $j \in I \cup H$, $c =(c_{i})_{i \in I} \in \{a,b,0\}^{I}$, and $\rho_{je}=\rho_{e}$ when $j=e \in H$. Hence, the cost share of a member is proportional to its GDP.

The simulation includes 146 countries for which the DI and bilateral trade data are available from 2006 to 2019. We take 5-year moving averages for all exogenous variables: DI, GDP, and the trade ratios. The endogenous variables are the club orientation choices of the US and China and the choices of non-superpowers regarding the club membership.\footnote{The US and China can choose any integer $x$ in range [0, 500] to place their clubs at $\frac{x}{500}$.} The simulation presents the equilibrium in each year. Hence,  rather than representing a historical timeline, the two stages in the game describe the situation in which the status-quo power has a first-mover advantage because of its established institutions.

The US is the status-quo power, who has a first-mover advantage through benchmarking and deterring. Therefore, allies, who have similar orientations with the status-quo power, are important. The DI is normalized to place countries in the range of [0,1], where the country with the highest democracy score is at 0 and the one with the lowest score is at 1.\footnote{We normalize the score $\ell_i$ for each country $i$ to be between 0 and 1 by setting $\ell_i = \frac{max_{DI} - DI_i} {max_{DI} - min_{DI}}$.} The US is placed around 0.2 and China around 0.8. 

Despite the first-mover advantage, the US is challenged at least in two dimensions today. First, the average trade ratio of countries with the US is declining, while that with China is increasing. In fact, China has surpassed the US as the dominant trading partner on average since 2013. Second, the GDP gap between the US and China is narrowing. In the following sections, we will examine the effects of those changes on the club orientation choices of the US and China, the club choices of all the other countries, and the resulting size of  clubs.

\subsection{Simulation from 2006 to 2019}

Let us first look at how changing trade ratios affect choices of non-superpowers. 
From 2006 to 2019, the average trade ratio of countries with the US decreased slightly from 0.13 to 0.10, while the ratio with China increased dramatically from 0.02  to 0.14. 

\begin{figure}[ht!]
	\begin{centering}		
		\subfigure[The choices of the US and China\label{fig:choice}]{\includegraphics[width=0.49\columnwidth]{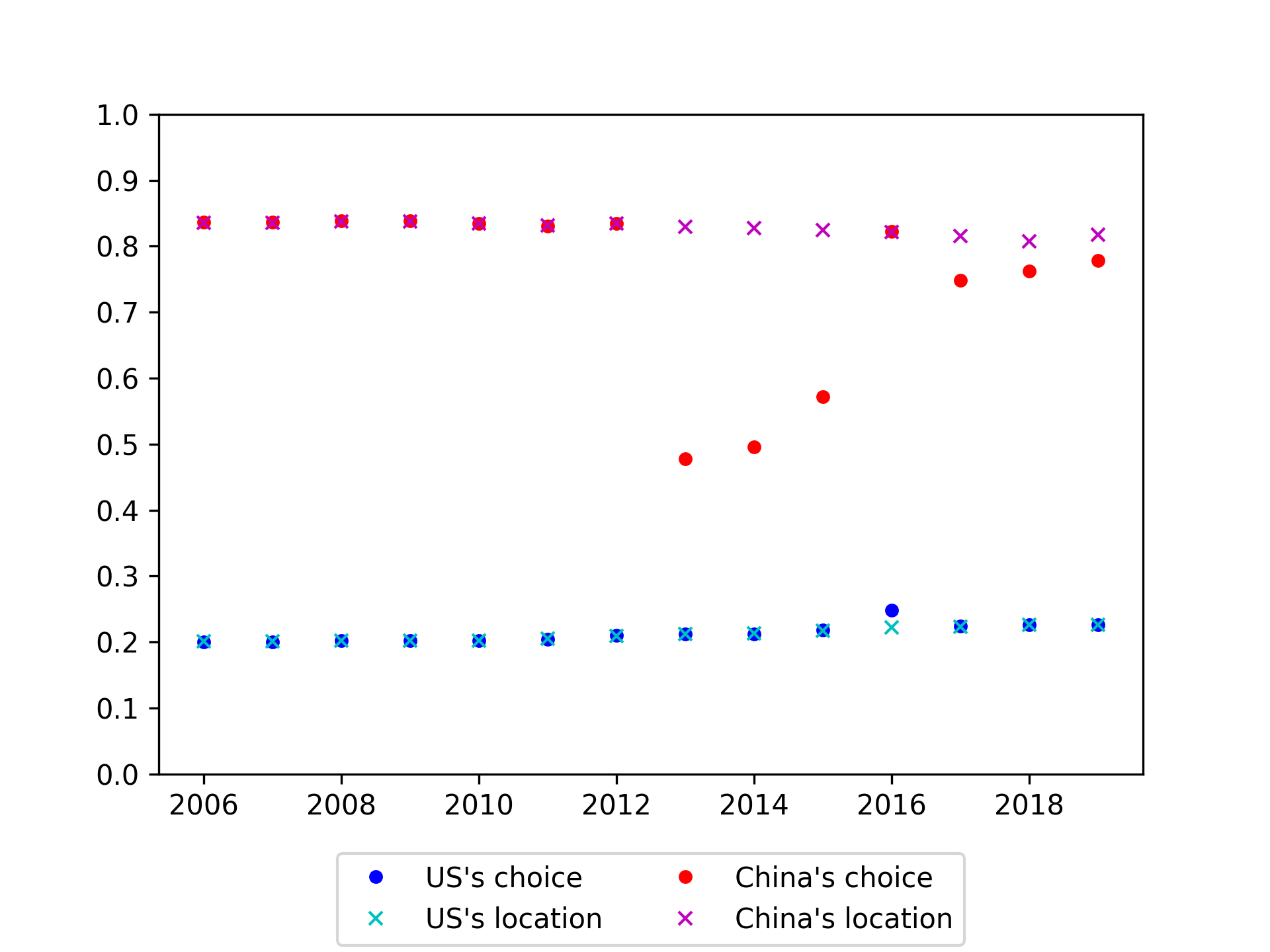}}
		\subfigure[Number of club members \label{fig:members}]{\includegraphics[width=0.49\columnwidth]{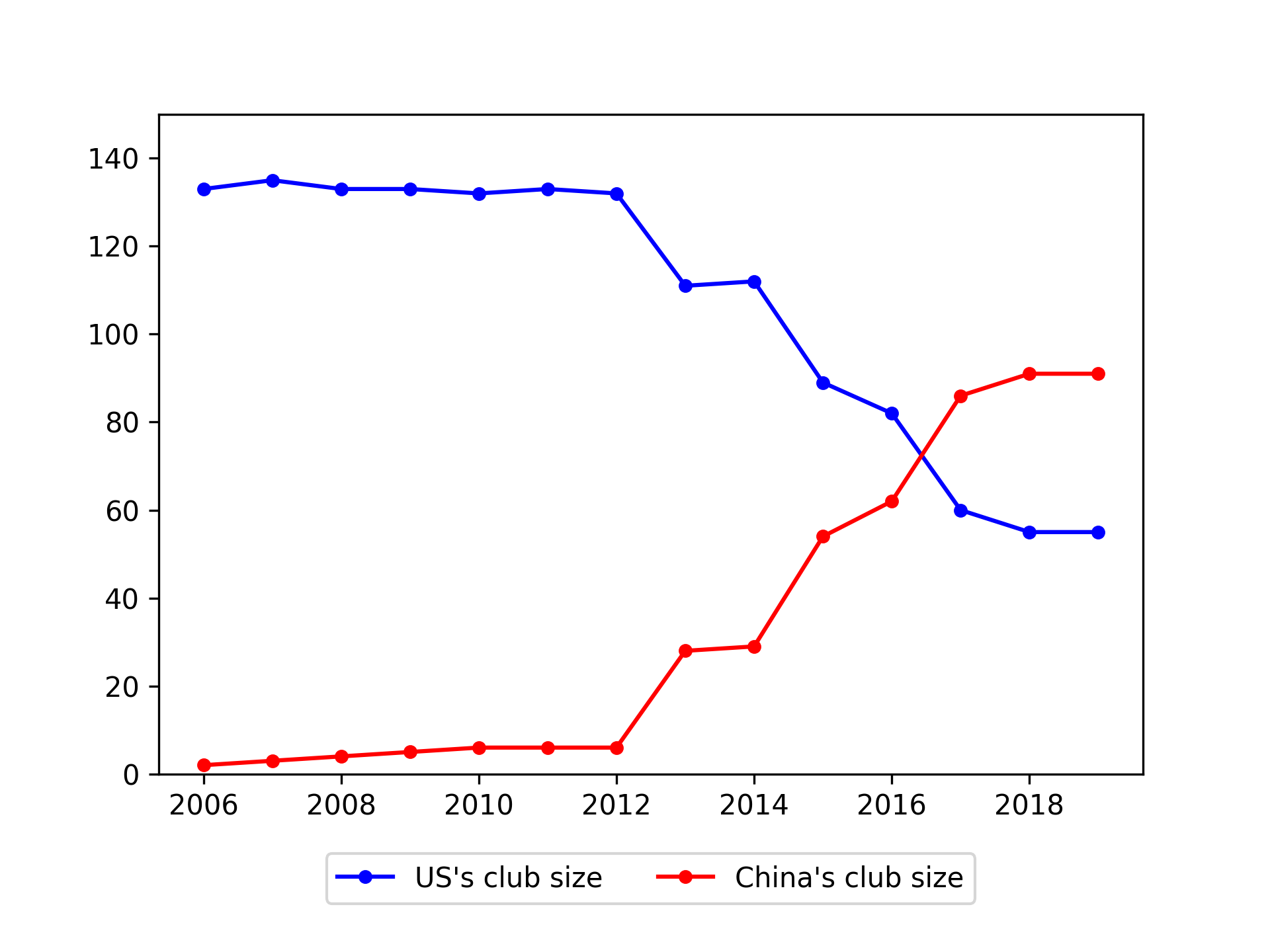}}
		\caption{The US and China with 144 countries}
	\end{centering}
\end{figure}

In 2006, when the majority of countries trade more with the US than China, most countries join the US's club even without the US compromising. When China chooses a centrist club orientation around 0.5 in 2013 (see Figure \ref{fig:choice}), it starts to gain significantly more members (see Figure \ref{fig:members}). In 2015, 59 out of 88 members in the US's club are robust club members (including Canada, Israel, Mexico, Peru, etc.), and 17 out of 53 members in China's club are robust club members (including Australia, Hong Kong, Laos, Mongolia, etc.). These member countries are highly dependent on the  superpower in terms of cost share and/or trade. 
Consistent with our theoretical result (Proposition \ref{dependency}), China starts to gradually compromise less as the robust club members become more dependent on it (see Figure \ref{fig:choice}).
In 2017, China gains more club members than the US thanks to its growing trade, although it chooses a less centrist orientation than before.

\begin{figure}[ht!]
	\begin{center}
		\subfigure[2006: $\sum_{i=1}^{n}g_{i,US}/n=0.13$ and $\sum_{i=1}^{n}g_{i,China}/n=0.02$]{\includegraphics[width=.9\columnwidth]{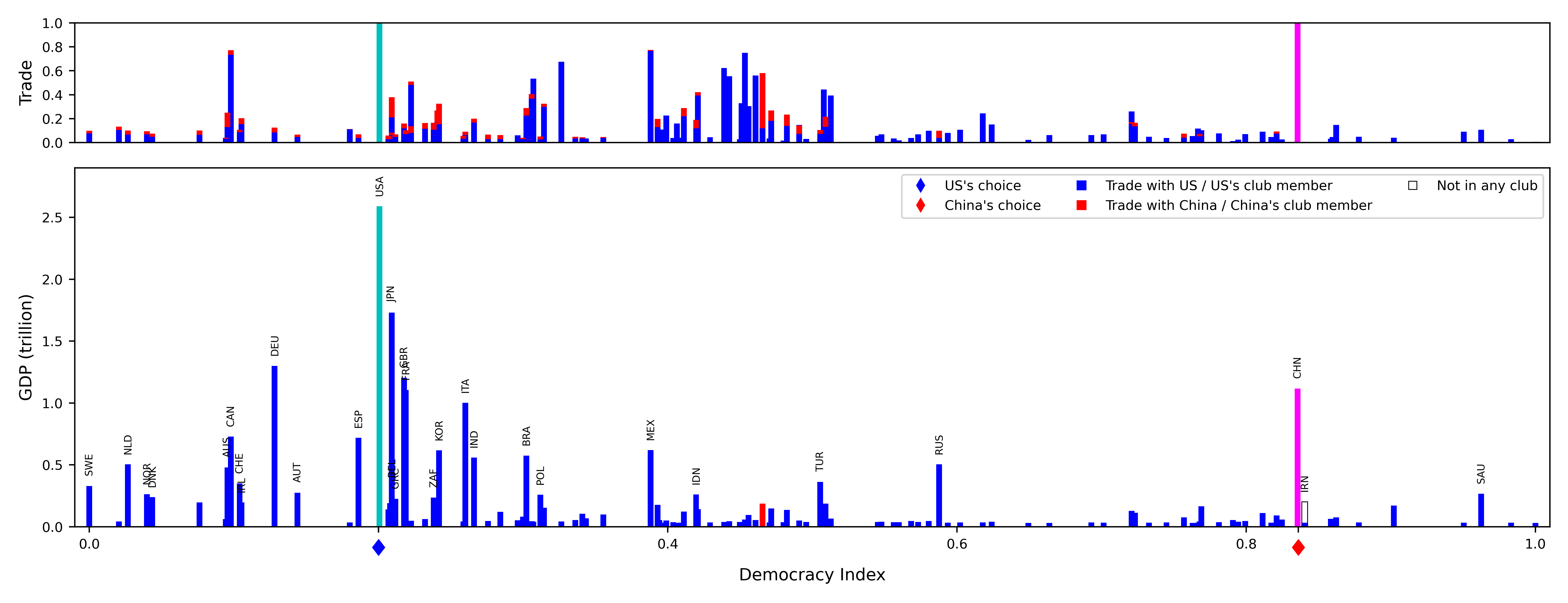}} \\
		\subfigure[2019: $\sum_{i=1}^{n}g_{i,US}/n=0.10$ and $\sum_{i=1}^{n}g_{i,China}/n=0.14$ ]{\includegraphics[width=.9\columnwidth]{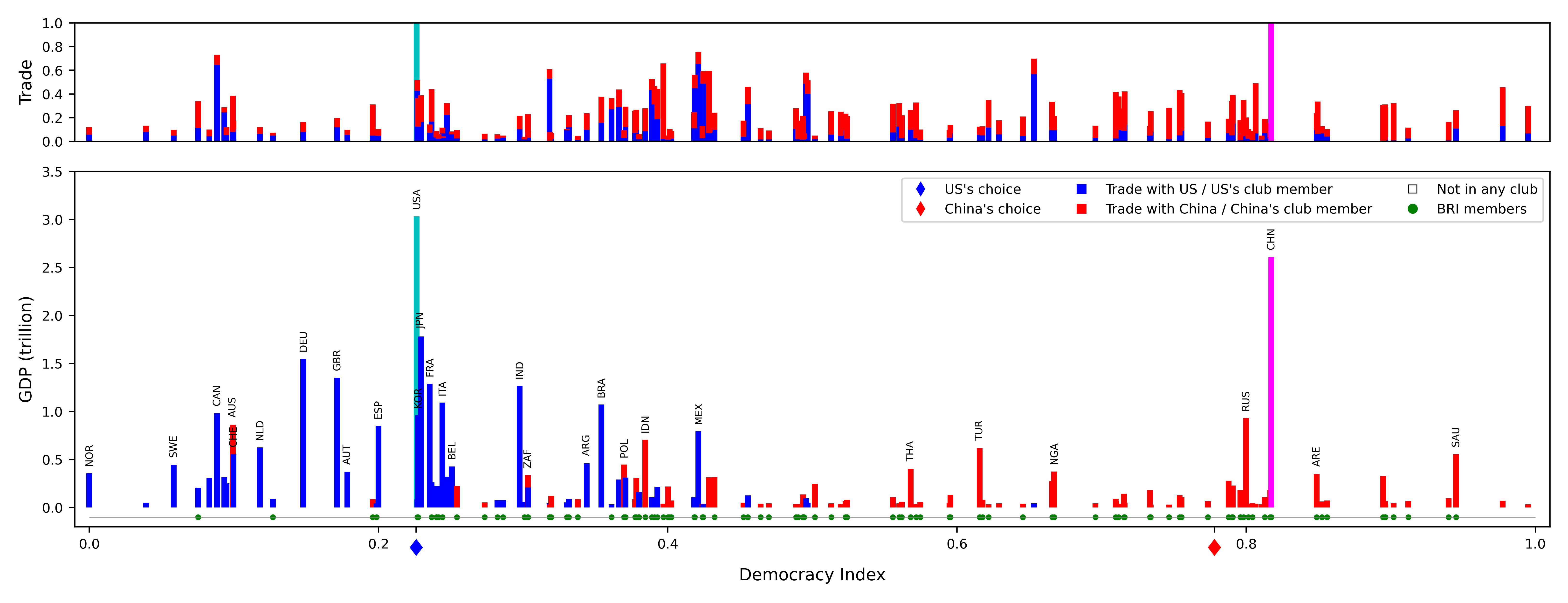}}
		\caption{The choices of the US and China and 144 countries }
		\label{fig:simulation_2007_2019}	
	\end{center}
	\floatfoot{
		\footnotesize\textit{Note}: For expositional purpose, the bar depicts $\ln(x+1.03)$, where $x$ is GDP in trillion current USD. The top 30 GDP countries are captioned. }
\end{figure}

Figure \ref{fig:simulation_2007_2019} shows each country's GDP and trade ratio with the US and China, its club choice---blue for the US's club and red for China's club, and the club orientation choices of the US and China by diamonds for 2006 and 2019. 
Regarding endowments, we observe that in 2006, GDP of the US was much higher than that of China, but the gap narrowed significantly  in 2019. The sum of GDP of China's club members is, however, still much smaller than that of the US's club members even in 2019.
Moreover, although the number of the US's club members has started falling  significantly since 2013, the GDP of its club members has not changed much. This implies that China attracts countries with a relatively low GDP and thus bears most of the cost to maintain the club through its growing GDP as Proposition \ref{endowment} predicts. 

\newpage

\subsection{Projection from 2030 to 2035}

This section presents simulation results based on the long-run GDP projection from 2030 to 2035 provided by the International Institute for Applied Systems Analysis \citep[see][]{riahi2017shared}.
On the other hand, no reliable projection for the DI and the trade ratio with the US and China is available.   
Therefore, we take the data for 2019 (5-year averages) for those variables to simulate the dynamics from 2030 to 2035.  

\begin{figure}[ht!]
	\begin{center}
		{\includegraphics[width=.9\columnwidth]{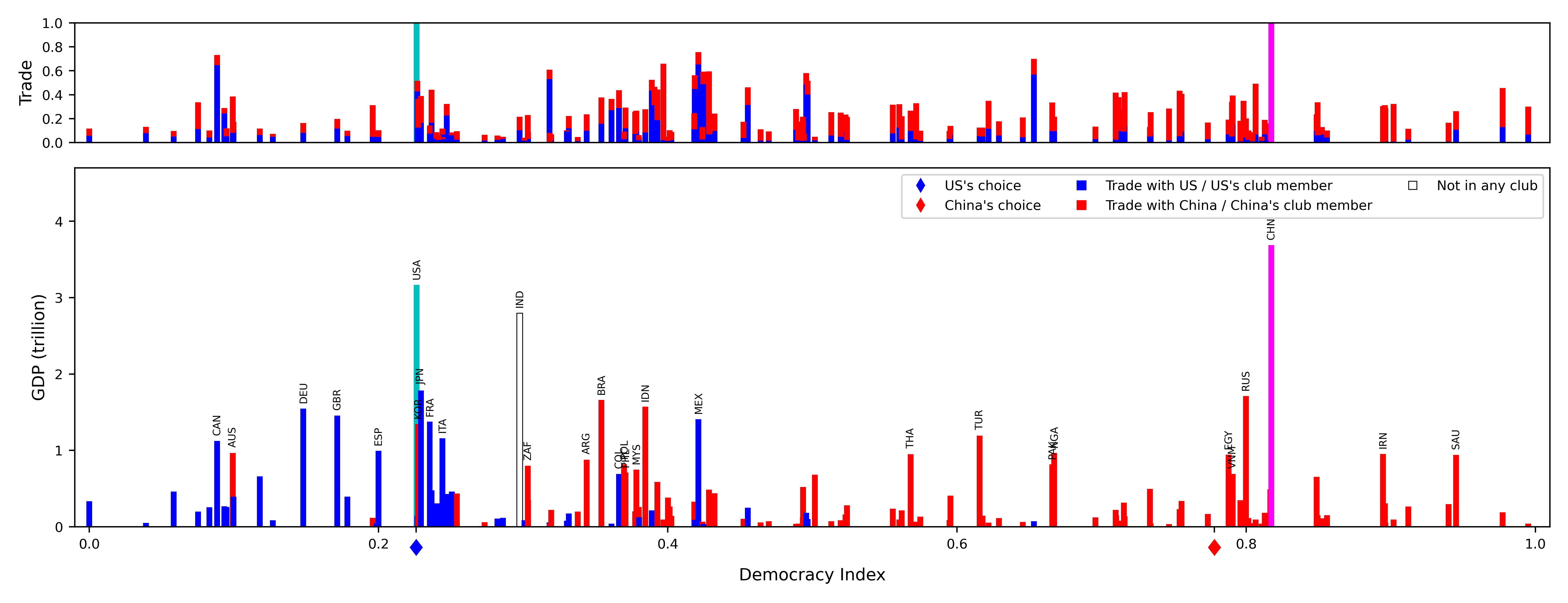}}
		\caption{China's club dominates the US's club in terms of club's total GDP by 2035.	\label{fig:simulation_2030_2035}}
	\end{center}
\end{figure}

Figure \ref{fig:simulation_2030_2035} shows that the US chooses a centrist club  orientation in an attempt to prevent the GDP of China's club members (\emph{excluding} China) from surpassing that of the US's club members (\emph{excluding} the US) in 2030.
However, by 2035, China's club starts to dominate the US's club even in terms of GDP.
Moreover, China chooses a centrist club orientation, making countries such as  Argentina,  Brazil, Korea, and the Philippines switch over to its club by 2035. 
The examples show that superpowers compromise when they compete over countries they would like to retain or obtain in their club. 
With India leaving the US's club, China's trade and GDP dominance out-powers the advantage of the US  having countries with a large GDP in the proximity in terms of orientation.
The major contributing factor for China's dominance  is India's departure from the US's club by 2030 as India's GDP and thus its cost share grow significantly as Proposition \ref{endowment2} predicts.

\section{Conclusion}\label{sec:conclusion}

We build a model where two superpowers play a Stackelberg game by choosing an orientation for their clubs in the presence of externalities. The cost each member pays depends on others who join the club, as members share the cost, while the benefit is decreasing in distance to the club. 
This presents a trade-off to the superpowers as they might want to attract more members by moving the club's orientation away from their own at the expense of their benefit from the club.  

The model presented in Section \ref{sec:two-stage} contains the specifications for simulations.  The simulation results in Section \ref{sec:US-China} highlight the behavior of superpowers and how non-superpowers shape the US-China competition for a sphere of influence as predicted by the comparative statics in Section \ref{sec:special}.

The simulation results capture China's strategic choice and growing sphere of influence between 2006 and 2019.  
For example, our simulation shows that China starts to take a more centrist orientation for its club in 2013, which coincides with the launch of the BRI, while expanding its sphere of influence.
The US's sphere of influence depends on countries with relatively large GDP and similar ideological orientations. On the other hand, China's sphere of influence depends on its growing trade share and GDP and the ideological compromise it makes to expand its club. The case of India described above highlights how even a single non-superpower might affect coalition formation in the presence of an externality and cause a shift in power balance between superpowers. 

\clearpage

\appendix
\newtheorem{lemmAA}{Lemma}
\renewcommand{\thelemmAA}{A.\arabic{lemmAA}}

\section{Preliminary results}

An \emph{abstract reduction system} is a pair $(X  , \rightarrow )$ where $X$ is a non-empty set and $ \rightarrow $ is a binary relation on $X$. An element $x  \in X$ is called an \emph{endpoint} in $(X, \rightarrow )$ iff there is no $x ^{ \prime } \in X $ such that $x  \rightarrow x ^{ \prime }$. We say that $\{x _{n} :n =0 ,1 ,\ldots \}$ (finite or infinite) in $X$ is a $ \rightarrow $\emph{-sequence} iff $x _{n} \rightarrow x _{n +1}$ (as far as $x _{n +1}$ is defined). We use $ \rightarrow ^{ \ast }$ to denote the reflexive and transitive closure of $ \rightarrow$. We say that $(X , \rightarrow )$ is \emph{weakly confluent} iff for each $x, y, z \in X $, if $x  \rightarrow y$ and $x  \rightarrow z$, then $y \rightarrow ^{ \ast } x ^{ \prime }$ and $z   \rightarrow ^{ \ast }x ^{ \prime}$ for some $x ^{ \prime } \in X$.

\begin{lemmAA}\label{NEW}
	\citep{newman1942theories} If an abstract reduction system $(X , \rightarrow )$ satisfies  \textbf{(N1)} each $ \rightarrow $-sequence is finite and \textbf{(N2)} $(X , \rightarrow )$ is weakly confluent, then for each $x  \in X $ there is a unique endpoint $x ^{ \prime } \in X $ such that $x  \rightarrow ^{ \ast }x ^{ \prime }$. 
\end{lemmAA}

The proof of following two lemmas can be seen in any textbook of discrete convex analysis, e.g., \cite{murota2003discrete}.

\section{Coalition formation}\label{sec:algorithm}

\subsection{Stage one}

Given $a$'s  choice $\ell_{a}$, the coalition-formation process in stage one yields a finite sequence $I_{1},...,I_{m}$ satisfying\footnote{The sequence is finite as the utility function \eqref{SEC} of non-superpowers is increasing in the club size.}
\begin{enumerate}
	\item {\em Join step-wise}: For each $k \in \{1,...,m\}, I_{k} \subseteq I$, and $I_{k} \cap I_{k^{\prime}}=\emptyset$ if $k \neq k^{\prime}$.
	
	\item {\em Everyone is better off}: For each  $k \in \{1,...,m\}$,
	\begin{equation}\label{attraction}
		g_{ia}(1-d_{i,\ell_{a}}) - \rho_{ia}(\cup_{t=1}^{k}I_{t}) \geq 0 \text{ for each }i \in I_{k}.
	\end{equation}
	
	\item {\em Maximality}: There is no nonempty subset $I_{m+1} \subseteq I \setminus \cup_{k=1}^{m}I_{k}$ such that \eqref{attraction} holds for $k = m+1$.
\end{enumerate}

By definition, the above process leads to a core outcome. 
Given $\ell_a$, multiple processes may satisfy the three conditions, yet we will show that the final coalition is uniquely determined regardless of the order of coalition formation. 
We have the following lemma.

\begin{lemmAA} \label{prop:1st_stage}
	Given $\ell_{a}$, the coalition-formation process in stage one is order-independent.
\end{lemmAA}

\begin{proof}[Proof of Lemma \ref{prop:1st_stage}]
	Let $X = 2^{I}$ and $E \rightarrow F$ if $E \subsetneq F$, all countries in $E$ are in $a$'s club (the club placed at $\ell_{a}$), and all countries in $F \setminus E$ obtain non-negative payoffs by joining $a$'s club together. Clearly, $(X, \rightarrow)$ describes a coalition-formation process and is a reduction system satisfying (N1) and (N2) in Lemma \ref{NEW}. Hence it is order-independent. Let $E^{*}$ be the final term of all coalition-formation processes. 
\end{proof}

\subsection{Stage two}

Given $\ell_{b}$ in stage two, the coalition-formation process in stage two yields a finite sequence $(\tilde{I}_{1}, I_{1}),...,(\tilde{I}_{p}, I_{p})$ satisfying

\begin{enumerate}
	\item {\em Shift step-wise}: For each $k \in \{1,...,p\}, \tilde{I}_{k} \subseteq I_{k} \subseteq I$, and $\tilde{I}_{k} \cap \tilde{I}_{k^{\prime}}=\emptyset$ if $k \neq k^{\prime}$.
	
	\item {\em Everyone is better off}: For each  $k \in \{1,...,p\}$, 
	\begin{itemize}
		\item[2.1] each $i \in \tilde{I}_{k}$ satisfies 
		\begin{equation}\label{shifting}
			\hspace{-0.43cm}
			\begin{array}{ll}
				g_{ib}(1-d_{i,\ell_{b}}) - \rho_{ib}(\cup_{t=1}^{k}\tilde{I}_{t}) \geq 0 & \text{if $i \not\in I^{*}(\ell_{a})$}\\
				g_{ia}(1-d_{i,\ell_{a}}) - \rho_{ia}(I^{*}(\ell_{a})\setminus\cup_{t=1}^{k-1}I_{t})< g_{ib}(1-d_{i,\ell_{b}}) - \rho_{ib}(\cup_{t=1}^{k}\tilde{I}_{t})  &\text{if $i \in I^{*}(\ell_{a})$}
			\end{array}
		\end{equation}	
		where $g_{ib}(1-d_{i,\ell_{b}}) - \rho_{ib}(\cup_{t=1}^{k}\tilde{I}_{t})   \geq 0$;
		
		\item[2.2] each $i \in I_{k}\setminus \tilde{I}_{k}$ satisfies  $g_{ia}(1-d_{i,\ell_{a}}) - \rho_{ia}(I^{*}(\ell_{a})\setminus\cup_{t=1}^{k-1}I_{t}) <0$ for  $i \in I^{*}(\ell_{a})$; 
		
	\end{itemize}
	
	\item[]	where we stipulate that $\cup_{t=1}^{0}I_{t}=\emptyset$.
	
	\item {\em Maximality}: There is no nonempty $\tilde{I}_{p+1} \subseteq I \setminus \cup_{k=1}^{p}\tilde{I}_{k}$ and $I_{p+1} \subseteq I \setminus \cup_{k=1}^{p}I_{k}$  such that the conditions in 2 hold.
\end{enumerate}

By definition, the above process leads to a core outcome. For each $k$, $\tilde{I}_k$ represents non-superpowers moving to $b$'s club, while $I_k$ denotes non-superpowers changing their behaviors from the previous sequence step. $\tilde{I}_k$ consists of two groups: countries that initially stayed out of $a$'s club but now join $b$'s club (first line of (\ref{shifting})), and countries that joined $a$'s club in stage one but now shift to $b$'s club (second line of (\ref{shifting})). For $I_k$, the first case overlaps with $\tilde{I}_k$’s second case, while the second case comprises countries that left $a$’s club in a previous step and now join $b$'s club.
Similar to stage one, we can show that the final coalition is unique. 
\begin{lemmAA}\label{prop:2nd_stage}	
	Given $\ell_{a}$, 	the coalition-formation process in stage two is order-independent for any $\ell_{b}$.
\end{lemmAA}

\begin{proof}[Proof of Lemma \ref{prop:2nd_stage}	]
	
	Fix  $\ell_{a}$ and  $\ell_{b}$. Let $X = \{0,a,b\}^{I}$. For each $c, c^{\prime} \in X$, we define $c\Delta c^{\prime} = \{i \in I: c_{i} \neq c^{\prime}_{i}\}$ and a binary relation $\rightarrow$ on $X$ as follows: For each $c, c^{\prime} \in X$, $c \rightarrow c^{\prime}$ iff (1) for each $i \in c\Delta c^{\prime} $, $u_{i}(c^{\prime}) \geq u_{i}(c)$ and (2) for some $i \in c\Delta c^{\prime} $, $u_{i}(c^{\prime}) > u_{i}(c)$. Define another binary relation $\Rightarrow$ on $\{0,a,b\}^{I}$ such that $c \Rightarrow c^{\prime}$ iff (1) $c \rightarrow c^{\prime}$ and (2) for each $i \in c\Delta c^{\prime}$, $c_{i} > c^{\prime}_{i}$ (here, we stipulate that $a >0>b$). It can be seen that $\Rightarrow$-sequences characterize the coalition-formation process in the second stage. 
	Now we show that when $I$ is finite, $(\{0,a,b\}^{I}, \Rightarrow)$ satisfies (N1) and (N2) in Lemma \ref{NEW}. First, it is clear that (N1) holds since every $\Rightarrow$-sequence is monotonic with respect to the club of $a$, i.e., for each $t \in \mathbf{N}$, $I_{a}(c^{t+1}) \subseteq I_{a}(c^{t})$. For (N2), let $c, c^{\prime}, c^{\prime\prime} \in \{0,a,b\}^{I}$ with $c \Rightarrow c^{\prime}$ and $c \Rightarrow c^{\prime\prime}$. Let $E^{\prime} = \{i \in c\Delta c^{\prime}: c^{\prime}_{i} = b\}$ and $E^{\prime\prime} = \{i \in c\Delta c^{\prime\prime}: c^{\prime\prime}_{i} = b\}$. It is clear that $E^{\prime} \cup E^{\prime\prime}$ can move as a whole to $b$ which increases everyone's payoff. For those who leaves $a$'s club and joins no club, the same argument holds. Hence, there is a $\hat{c}$ such that $c^{\prime} \Rightarrow \hat{c}$ and $c^{\prime\prime} \Rightarrow \hat{c}$.  Therefore, (N2) is also satisfied.  By Lemma \ref{NEW}, the process is order-independent.
\end{proof}

For each disjoint $E,F \subseteq I$, we define $c^{\langle E,F\rangle}\in \{a,b,0\}^{I}$ such that the $i$-th entry is $a$ if $i \in E$, $b$ if $i \in F$, and $0$ if $i \in I\setminus (E \cup F)$. 
Given $\ell_a$ and $\ell_b$, we define $I_{a}(\ell_{a},\ell_{b}) = I^{*}(\ell_{a})\setminus (\cup_{k=1}^{p}I_{k})$ and $I_{b}(\ell_{a},\ell_{b}) = \cup_{k=1}^{p}\tilde{I}_{k}$. 
Lemma \ref{prop:2nd_stage} guarantees that $I_{a}(\ell_{a},\ell_{b})$ and $I_{b}(\ell_{a},\ell_{b})$ are well defined and stable. 
At the end of the game, $a$'s club is $I_{a}(\ell_{a},\ell_{b})\cup \{a\}$ and $b$'s club is $I_{b}(\ell_{a},\ell_{b}) \cup \{b\}$, which, according to \eqref{UYT}, determine $a$ and $b$'s payoff. 

\section{Main proofs}

\begin{proof}[Proof of Lemma \ref{lemma_spe}]
By the way of backward induction we start from $b$'s choice. Given $a$'s choice,
suppose a tie (of a maximum payoff) between $b$'s choices. There
are two possibilities. First, $b$ might be indifferent between multiple orientations.
There must be an orientation closest to $b$'s, and $b$ chooses the closest one by Assumption \ref{assumption1}. Second, $b$ might be indifferent between some orientation(s) and not establishing a club. Then, $b$ chooses not to establish a club by \ref{assumption2}. Analogously, $a$'s choice is unique. Hence, the result in Lemma \ref{lemma_spe} holds.
\end{proof}

\begin{proof}[Proof of Proposition \ref{dependency}]	
	Suppose that in game $G^1$, $a$ and $b$ chooses $(\ell_a^1, \ell_b^1)$ and forms two clubs in the SPE outcome. 
	First, consider another game $G^2$ where $g_{ib}^2 \geq g_{ib}^1$ for all $i \in \underline{I}_b (\ell_a^1) \neq \emptyset$ with at least one strict inequality. Then, in $G^2$, if $a$ keeps choosing $\ell_a^1$, $b$ attracts the same club with weakly less compromise.
	Since all non-superpowers not in $\underline{I}_b (\ell_a^1)$ do not change their behavior, given $\ell_a=\ell_a^1$, compromising more leads to the same outcome for $b$ as in $G^1$. Hence, $b$'s best response is to compromise weakly less in this subgame. In other subgames, $b$ either compromises more and attracts more members in $G^2$ than in $G^1$, or compromises weaky less without expanding its club. In subgames where $b$ compromises more, $a$'s utility weakly decreases and thus, $\ell_a^1$ remains optimal for $a$ among these subgames. In subgames where $b$ compromises weakly less, $a$'s utility weakly increases. Therefore, in the SPE outcome, $b$ compromises weakly less.
	
	Second, consider another game $G^3$ where $g_{ia}^3 \geq g_{ia}^1$ for all $i \in \underline{I}_a (\ell_a^1) \neq \emptyset$ with at least one strict inequality. Since all non-superpowers not in $\underline{I}_a (\ell_a^1)$ do not change their required minimum club endowment to join $a$'s club, if $a$ keeps choosing $\ell_a^1$ or compromises more in $G^3$, there is no change in $b$'s best response. Besides, it may be beneficial for $a$ to compromise less due to the non-superpowers becoming more dependent on it and thus a weakly larger club size than in $G^1$. Therefore, in the SPE outcome, $a$ compromises weakly less.
\end{proof}

\begin{proof}[Proof of Proposition \ref{endowment}] Since $\ell_a$ and $\ell_b$ are fixed, the benefit $g_{ie} (1-d_{ie})$ remains unchanged for all $i \in I$ and for all $e \in \{a,b\}$.
	
	Part 1. Since $\rho_{ie}(I_e;\omega')=\rho_{ie}(I_e;\omega)$ for all $i \in I_e \cup \{e\}$, all existing members remain in $e$'s club. Since $\omega'_i > \omega_i$ for all $i \in I_e \cup \{e\}$, the cost of joining $e$'s club falls for all others, i.e., $\rho_{je}(I_e \cup J;\omega') < \rho_{je}(I_e \cup J;\omega)$ for all $j \in J$ and for all $J \subseteq I \backslash I_e$. Therefore, $I'_e \supseteq I_e$. 
	
	Part 2. Since $\omega'_e > \omega_e$, the cost of joining $e$'s club falls for all non-superpowers, i.e., $\rho_{je}(J;\omega') < \rho_{je}(J;\omega)$ for all $j \in J$ and for all $J \subseteq I$. Therefore, $I'_e \supseteq I_e$. 
\end{proof}

\bibliographystyle{ecta}
\bibliography{hegemon} 

\end{document}